\documentclass[a4,12pt]{article}

\usepackage{colortbl}
\usepackage{ascmac}
\usepackage{bm}
\usepackage[dvipdfmx]{graphicx}
\usepackage{amsmath}
\usepackage{amssymb}
\usepackage{amsfonts}
\usepackage{mathrsfs}
\usepackage{multirow}

\newcommand{\ol}{\overline}

\newcommand{\wt}{\widetilde}

\newcommand{\CC}{\mathbb{C}}
\newcommand{\ZZ}{\mathbb{Z}}

\newcommand{\RR}{\mathbb{R}}

\def\tr{\mathop{\rm tr}\nolimits}
\def\Tr{\mathop{\rm Tr}\nolimits}
\def\diag{\mathop{\rm diag}\nolimits}

\def\Pexp{\mathop{\rm Pexp}\nolimits}

\newcommand{\rap}[2]
{\setbox1=\hbox{#1}%
\setbox2=\hbox to\wd1{\hss #2\hss}%
\mbox{\rlap{\box1}\box2}}

\title{Orbifold Schur index and IR formula}

\begin{document}

\begin{titlepage}
\title{
\begin{flushright}
\normalsize{
TIT/HEP-661\\
October 2017}
\end{flushright}
       \vspace{1cm}
       Orbifold Schur index and IR formula
       \vspace{1cm}}
\author{
Yosuke Imamura\thanks{E-mail: \tt imamura@phys.titech.ac.jp}$^{~1}$
\\[30pt]
{\it $^1$ Department of Physics, Tokyo Institute of Technology,}\\
{\it Tokyo 152-8551, Japan}
}
\date{}

\thispagestyle{empty}

\vspace{0cm}

\maketitle
\begin{abstract}
We discuss an orbifold version of the Schur index defined as the supersymmetric partition function
in $\bm{S}^3/\ZZ_n\times\bm{S}^1$.
We first give a general formula for Lagrangian theories obtained by localization technique,
and then suggest a generalization of the Cordova and Shao's IR formula.
We confirm the generalized IR formula gives the correct answer
for systems with free hypermultiplets if we tune the background fields so that they are
invariant under the orbifold action.
Unfortunately, we find disagreement for theories with dynamical vector multiplets.
\end{abstract}
\end{titlepage}

\section{Introduction}
Exactly calculable quantities in supersymmetric field theories
play important roles in recent progress in quantum field theories.
The Schur index of ${\cal N}=2$ superconformal field theories
is one of such quantities.
There are different ways to calculate the index.

The Schur index is a specialization of the superconformal index of
${\cal N}=1$ superconformal field theories \cite{Kinney:2005ej,Romelsberger:2005eg}.
(See \cite{Gadde:2011uv} for various limits of the superconformal index.)
The superconformal index
can be regarded as the supersymmetric partition function
in $\bm{S}^3\times\bm{S}^1$.
For Lagrangian theories
it is defined as the path integral in the background,
and we can reduce it as a finite-dimensional
integral by using localization method.
This is also available for a non-Lagrangian theory if we know a
UV Lagrangian theory that flows to the theory.

Cordova and Shao
\cite{Cordova:2015nma} 
proposed an interesting formula for the Schur index
which gives the index as
the trace of so-called a quantum monodromy operator.
With this formula we can calculate
the index from the information of the
BPS spectrum of the theory in
the Coulomb branch.
Although the BPS spectrum
depends on the Coulomb moduli parameters
and jumps on walls of marginal stability
the quantum monodromy operator is wall-crossing invariant, and
so is the index.
The formula is generalized to decorated indices by introducing defect operators\cite{Cordova:2016uwk,Cordova:2017ohl,Cordova:2017mhb}.

For class S theories, which are realized on M5-branes wrapped on Riemann surfaces,
the Schur index is expressed as a correlation function of a two-dimensional
topological field theory
on the associated Riemann surface \cite{Gadde:2011ik,Gadde:2009kb,Buican:2015ina,Buican:2015tda,Song:2015wta,Buican:2017uka,Song:2017oew}.

The Schur index receives contributions
of a special class of gauge invariant operators, which are
called Schur operators.
Beem et. al. \cite{Beem:2013sza}
shows that
the set of Schur operators
form a chiral algebra of two-dimensional CFT.
Once we identify the chiral algebra associated with a four dimensional
theory the Schur index is obtained as the
vacuum character of the chiral algebra.
The characters of other modules are also important and
are related to line and surface operator insertions \cite{Cecotti:2015lab,Cordova:2017mhb}.
For theories of class S there is a prescription to obtain the corresponding
chiral algebra \cite{Beem:2014rza}.

The purpose of this paper is investigate a generalization of the Schur index
defined by replacing the background manifold $\bm{S}^3\times\bm{S}^1$
by its orbifold $\bm{S}^3/\ZZ_n\times\bm{S}^1$.
Such an orbifold generalization
for the superconformal index of ${\cal N}=1$ and ${\cal N}=2$ theories
has already been
investigated in
\cite{Benini:2011nc,Alday:2013rs,Razamat:2013jxa,Razamat:2013opa,Kels:2017toi}.
The index is defined in \cite{Benini:2011nc} by using $\ZZ_n$
which preserves a supercharge of a specific chirality.
Because we need only one supercharge (and its hermitian conjugate) for the definition of the
superconformal index this orbifolding is consistent with the
definition of the index.
However, in the case of Schur index, we need to preserve
two supercharges with opposite chirality (and their hermitian conjugate).
Therefore, we need to modify the definition of the orbifold index by
introducing extra $SU(2)_R\times U(1)_r$ twist.
The application of the localization technique for the orbifold Schur index is
straightforward
and will be shown in Section \ref{UV.sec}.
Then, we discuss a generalization of the IR formula
to the orbifold case.
We suggest a natural generalization of the Cordova and Shao's
formula
based on a physical interpretation of the formula,
and apply it to some simple examples.
For systems consisting of free hypermultiplets we
find agreement of the UV and IR formulae.
Unfortunately, however, for more general systems including dynamical
vector multiplets the suggested formula does not give the desired results.

\section{UV formula}\label{UV.sec}
Let us consider an ${\cal N}=2$ superconformal field theory
defined in $\bm{S}^3\times\RR_t$.
Let $H$, $J$, $\ol J$, $R$, and $r$ be
the Hamiltonian, the third component of the left-handed spin,
that of the right-handed spin,
the $SU(2)_R$ Cartan generator, and the $U(1)_r$ charge, respectively.
The superconformal index is defined by
\begin{align}
I(p,q,t,\vec z_F)=\Tr\left[e^{2\pi i(J+\ol J)}
x^{\mu_x}
p^{\mu_p}
q^{\mu_q}
t^{\mu_t}
\vec z_F^{\vec T_F}
\right],
\label{defsci}
\end{align}
where
the trace is taken over all gauge invariant states in $\bm{S}^3$.
We denote the set of Cartan generators of the flavor group $F$ by
$\vec T_F=(T_{F,1},\ldots,T_{F,r_F})$.
Other Cartan generators are defined by
\begin{align}
\mu_x&
=\frac{1}{2}H-J-R+\frac{1}{4}r
=\{Q,Q^\dagger\}
,\\
\mu_p&
=\frac{1}{2}H-\ol J-R-\frac{1}{4}r
=\{\ol Q,\ol Q^\dagger\}
,\\
\mu_q&=J+\ol J+R,\\
\mu_t&=R+\frac{1}{2}r,
\label{danddbar}
\end{align}
where $Q$ and $\ol Q$ are supercharges with the following quantum numbers.
\begin{align}
Q:(H,J,\ol J,R,r)&=(\tfrac{1}{2},-\tfrac{1}{2},0,+\tfrac{1}{2},-1),\nonumber\\
\ol Q:(H,J,\ol J,R,r)&=(\tfrac{1}{2},0,-\tfrac{1}{2},+\tfrac{1}{2},+1).
\end{align}
The definition (\ref{defsci}) respects $Q$.
Namely, all Cartan generators used in (\ref{defsci})
commute with $Q$.
Because $\mu_x$ is $Q$-exact
the index (\ref{defsci}) is independent of $x$.%
\footnote{Our convention is different from the standard one used, for example, in \cite{Gadde:2011uv}.
The standard one is obtained from ours by
the replacement $t\rightarrow t/q$, and then the Schur limit is given by $t\rightarrow q$,
while in our convention the Schur limit is $t\rightarrow 1$.}

The superconformal index can be regarded as the
supersymmetric partition function in $\bm{S}^3\times\bm{S}^1$.
If the theory has a Lagrangian description
with gauge group $G$ we can define
the index as the path integral.
By using localization technique
we can reduce the path integral
to the finite dimensional integral
\begin{align}
I=\int d\mu
\Pexp i,
\label{ioint}
\end{align}
where $\Pexp$ is the plethystic exponential defined by
\begin{align}
\Pexp f(x_1,x_2,,\ldots,x_k)
=\exp\sum_{m=1}^\infty\frac{1}{m} f(x_1^m,x_2^m,,\ldots,x_k^m).
\end{align}
$i$ is the one-particle index defined by
\begin{align}
i(p,q,t,\vec z)=\tr\left[e^{2\pi i(J+\ol J)}
x^{\mu_x}
p^{\mu_p}
q^{\mu_q}
t^{\mu_t}
\vec z^{\vec T}
\right],
\label{defopi}
\end{align}
where the trace is taken over all one-particle states
including gauge non-invariant states.
$\vec T$ includes both flavor and gauge Cartan generators,
and $\vec z$ is the set of the corresponding fugacities.
Namely, $\vec z^{\vec T}=\vec z_F^{\vec T_F}\vec z_G^{\vec T_G}$.
$\int d\mu$ is the integral over the
gauge fugacities
\begin{align}
\int d\mu=\frac{1}{|W_G|}\prod_{i=1}^{r_G}\oint\frac{dz_{G,i}}{2\pi iz_{G,i}},
\end{align}
where $|W_G|$ is the size of the Weyl group of $G$.

The Schur index is obtained from the superconformal index by the specialization
$t=1$.
Then $\mu_t$ disappears from (\ref{danddbar}),
and all the remaining Cartan charges commute with
not only $Q$ but also $\ol Q$.
Therefore, the Schur index receives contributions of
operators that carry $\mu_x=\mu_p=0$.
Such operators are called Schur operators.
As a result, the Schur index is a function of the superconformal fugacity $q$ and
the flavor fugacities $\vec z_F$.

There is another index that receives from the same set of operators as the Schur index.
It is the Macdonald index defined from the superconformal index by taking the limit $p\rightarrow0$.
We do not consider the Macdonald index here because its definition does not respect the
supercharges $Q$ and $\ol Q$ which are important when we discuss BPS configuration associated with the IR formula.
Even so, the Macdonald index is closely related to the Schur index, and can be reproduced from the
chiral algebra up to some ambiguity that can be fixed by using information of 4d SCFT \cite{Song:2016yfd}.
In the recent paper \cite{fluder} an orbifold version of the Macdonald index is studied,
and a close relation to the chiral algebra is observed.
It would be important to study relation between the orbifold Schur index defined below and the
orbifold Macdonald index in \cite{fluder}.

The one-particle Schur index is given by
\begin{align}
i=r_Gi_{U(1)}+i_V+i_H,
\end{align}
where $i_{U(1)}$, $i_V$, and $i_H$ represent
the contributions of a single $U(1)$ vector multiplet,
charged vector multiplets, and half-hypermultiplets, respectively.
They are explicitly given as
\begin{align}
i_{U(1)}=\frac{-2q}{1-q},\quad
i_V=-\frac{1+q}{1-q}\chi'_{\rm adj}(\vec z_G),\quad
i_H=\frac{q^{\frac{1}{2}}}{1-q}\chi_{R_H}(\vec z_F,\vec z_G),
\label{eq11}
\end{align}
where
$\chi_{\rm adj}'\equiv\chi_{\rm adj}-r_G$ is the character of the adjoint representation
of $G$
with the Cartan contribution subtracted,
and $\chi_{R_H}$ is the character of the $G\times F$ representation of the half-hypermultiplets.%
\footnote{The reason why $i_V$ in (\ref{eq11}) is not simply $i_{U(1)}\chi_{\rm adj}'$ but $(i_{U(1)}-1)\chi_{\rm adj}'$
is that we include the measure factor (Vandermonde determinant) in $i_V$.
In other words ``$-1$'' is the contribution of the Faddev-Popov ghost associated with
constant gauge transformation over ${\bm S}^3$.}
Because $i_{U(1)}$ does not depend on the gauge fugacities $\vec z_G$
we can factor out the Cartan contribution
in (\ref{ioint}) as $\lambda^{r_G}$, with $\lambda$ defined by
\begin{align}
\lambda=
\Pexp i_{U(1)}=(q;q)_\infty^2,
\label{u1shur}
\end{align}
where $(z;q)_\infty$ is the $q$-Pochhammer symbol defined by
\begin{align}
(z;q)_\infty=\prod_{k=0}^\infty(1-zq^k).
\end{align}

A generalization of the superconformal index to the orbifold background $\bm{S}^3/\ZZ_n\times\bm{S}^1$
was first investigated in \cite{Benini:2011nc}.
They defined the orbifold with the discrete group $\ZZ_n$ generated by
\begin{align}
\ol g_n=\omega_n^{2\ol J},\quad
\omega_n\equiv e^{\frac{2\pi i}{n}}.
\end{align}
This is consistent to the definition (\ref{defsci})
of the superconformal index
in the sense that
$\ol g_n$ commutes with the supercharge $Q$.
If the theory has gauge and/or flavor symmetry
we can turn on holonomies $\vec h$ by using
$\ol g_n\omega_n^{\vec h\vec T}$ instead of $\ol g_n$ as the $\ZZ_n$ generator.
Once we choose the $\ZZ_n$ generator, it is straightforward to
generalize the formula (\ref{ioint}) to the orbifold case.
What we have to do first is projecting away the contribution of
$\ZZ_n$ non-invariant states from
the one-particle index (\ref{defopi}).
This projection is carried out by inserting the projection
\begin{align}
\frac{1}{n}\sum_{k=0}^{n-1}(\ol g_n\omega_n^{\vec h\vec T})^k
\end{align}
into the trace in (\ref{defopi}).
Because
$\ol g_n$ can be expressed in terms of the Cartan generators
appearing in (\ref{defsci}) as
$\ol g_n=\omega_n^{2\ol J}=\omega_n^{\mu_x-\mu_p+\mu_q-\mu_t}$,
insertion of $(\ol g_n\omega_n^{\vec h\vec T})^k$ is equivalent to the replacement of the fugacities
\begin{align}
(x,p,q,t,\vec z)\rightarrow
(\omega_n^kx,\omega_n^{-k}p,\omega_n^kq,\omega_n^{-k}t,\omega_n^{k\vec h}\vec z),
\label{vartor}
\end{align}
and the one-particle index for the $\ZZ_n$ orbifold is
\begin{align}
i_n^{\vec h}(p,q,t,\vec z)=
\frac{1}{n}\sum_{k=0}^{n-1}
i(\omega_n^{-k}p,\omega_n^kq,\omega_n^{-k}t,\omega_n^{k\vec h}\vec z).
\end{align}
By using this orbifold one-particle index, the orbifold index is given by
\begin{align}
I_n^{\vec h_F}=\sum_{\vec h_G} e^{\varepsilon(\vec h)}\int d\mu\Pexp i_n^{\vec h},
\label{inh}
\end{align}
where
the factor $e^{\varepsilon(\vec h)}$ is the contribution of zero-point energy.
For ordinary index this factor just gives an overall factor and
usually neglected.
However, in the orbifold case this factor is important
because it depends on the gauge holonomy $\vec h_G$.
The zero-point factor
$e^{\varepsilon(\vec h)}$ is easily obtained by taking the product of all contributions
of one-particle states.
For example, if the one-particle index is expanded as
$i_n^{\vec h}=\sum_ic_ip^{a_i}q^{b_i}(\cdots)$ (We explicitly show $p$ and $q$ dependence
for simplicity, and the dots include other fugacities.)
then
the corresponding zero-point factor is given by
$e^{\varepsilon(\vec h)}=\prod_i(p^{\frac{a_i}{2}}q^{\frac{b_i}{2}}\cdots)^{c_i}$.
Although this infinite product is usually divergent,
we can obtain finite result
by using an appropriate regularization such as $\zeta$-function regularization.

Now, let us apply the same prescription of the orbifolding to
the Schur index.
We cannot obtain
the orbifold Schur index by the specialization $t\rightarrow 1$
from the orbifold superconformal index above
because
this contradicts the orbifold action (\ref{vartor}).
For the consistency to $t=1$
we use
\begin{align}
g_n=
\ol g_n e^{\frac{2\pi i}{n}(R+\frac{1}{2}r)}
=e^{\frac{2\pi i}{n}(\mu_x-\mu_p+\mu_q)},
\label{gdef}
\end{align}
instead of $\ol g_n$.
Namely, we combine $\ol g_n$ with the additional $SU(2)_R\times U(1)$ twist
to keep both $Q$ and $\ol Q$ invariant.
Again, we can turn on holonomies $\vec h$, and
the insertion of $(g_n\omega_n^{\vec h\vec T})^k$ is equivalent to the replacement,
\begin{align}
(x,p,q,\vec z)
\rightarrow(\omega_n^k x,\omega_n^{-k}p,\omega_n^k q,\omega_n^{k\vec h}\vec z).
\end{align}
Therefore, the orbifold one-particle index is
\begin{align}
i_n^{\vec h}(q,\vec z)=\frac{1}{n}\sum_{k=0}^{n-1}i(\omega^kq,\omega^{k\vec h}\vec z).
\end{align}
With this one-particle index
we can calculate the orbifold Schur index
of Lagrangian theories by using
(\ref{inh}), which we call UV formula in the following.

Before discussing the IR formula for the orbifold Schur index,
let us apply the UV formula to some simple systems.
In the following examples the zero-point factors give overall factors independent
of the holonomies and we omit them.

\paragraph{$U(1)$ vector multiplet}

The $U(1)$ contribution in $\bm{S}^3$ is given by $\lambda$ in (\ref{u1shur}).
The $\ZZ_n$ action is simple phase rotation of variable $q$,
and we obtain the orbifold one-particle index
$i_n=-2q^n/(1-q^n)$.
The orbifold Schur index is
\begin{align}
\lambda_n=
\Pexp\left(\frac{-2q^n}{1-q^n}\right)=(q^n;q^n)_\infty^2.
\label{lambdan}
\end{align}

\paragraph{Free hypermultiplet}
Let us consider the system of a single free hypermultiplet $(q,\wt q)$.
This has $SU(2)_F$ flavor symmetry.
Let $z$ be the fugacity for $U(1)_F\subset SU(2)_F$
and $F$ be the corresponding generator such that
$q$ and $\wt q$ carry $F=+1$ and $-1$, respectively.
The one-particle index before $\ZZ_n$ projection is
\begin{align}
i(q,z)=\frac{q^{\frac{1}{2}}}{1-q}(z+z^{-1}).
\end{align}
The $\ZZ_n$ generator acts on the variables as
\begin{align}
g_n\omega_n^{hF}:(q,z)\rightarrow(\omega_n q,\omega_n^hz).
\end{align}
By definition the generator $g_n\omega_n^{hF}$ must satisfy $(g_n\omega_n^{hF})^n=1$.
(Otherwise the fiber bundle associated with the fields is ill-defined.)
Due to the fractional $SU(2)_R$ charge
$R=-1/2$ of $q$ and $\wt q$,
$g_n^n$ acts on the hypermultiplet as $-1$.
To compensate this,
we need to take fractional holonomy $h\in\ZZ+\frac{1}{2}$.
Then, after the projection we obtain
\begin{align}
i_n^h(q,z)=\frac{q^{\frac{1}{2}}}{1-q^n}
(q^{[-h-\frac{1}{2}]_n}z+q^{[h-\frac{1}{2}]_n}z^{-1}),\quad
h=\frac{1}{2},\ldots,n-\frac{1}{2},
\label{hyperzn}
\end{align}
where $[x]_n$ for $x\in\ZZ$ is the minimum non-negative integer satisfying
$[x]_n\equiv x\mod n$.
In the $\ZZ_2$ case, the one-particle indices for two holonomies
$h=\pm1/2$ are
\begin{align}
i_2^{+\frac{1}{2}}=\frac{q^{\frac{1}{2}}}{1-q^2}(qz+z^{-1}),\quad
i_2^{-\frac{1}{2}}=\frac{q^{\frac{1}{2}}}{1-q^2}(z+qz^{-1}).
\label{hyperz2}
\end{align}

\paragraph{QED}
Let us consider $U(1)$ gauge theory with $N_f$ hypermultiplets.
This theory is not conformal, but we can obtain the index by applying the localization formula.
When we consider orbifold index, we should be careful about the fact that $U(1)_r$
is broken to the subgroup $\ZZ_{N_f}\subset U(1)_r$ by anomaly.
The orbifold action must be consistent to this unbroken symmetry.
For example, in the case of $\ZZ_2$ orbifold, $N_f$ must be even.
For $N_f=2$ the $\ZZ_2$ orbifold index is given by
\begin{align}
I^{\frac{1}{2}}_2
&=
\lambda_2
\sum_{h=\pm\frac{1}{2}}\oint\frac{dz}{2\pi iz}
\Pexp\left(2i_2^h\right)
=2(1+2q^2+8q^4+\cdots),
\label{qednf2}
\end{align}
where $i_2^h$ is the one-particle index (\ref{hyperz2})
for a single hypermultiplet.
We did not turn on the $SU(N_F)$ flavor symmetry for simplicity,
and we omitted the zero-point factor which does not depend on $h=\pm1/2$.

\section{IR formula}
The IR formula proposed by Cordova and Shao \cite{Cordova:2015nma}
gives the Schur index by using the information of BPS spectrum
in the Coulomb branch.

Let $\Gamma$ be the charge lattice
of flavor and gauge charges, and
$\langle\gamma,\gamma'\rangle$ be the associated Dirac pairing.
The central charge of a particle is determined by its charge $\gamma\in\Gamma$,
and we denote it by $Z_\gamma$.
The flavor sublattice $\Gamma_F\subset\Gamma$ is defined as the set of
charges $\gamma$ which have vanishing pairing
$\langle\gamma,\gamma'\rangle=0$ with
arbitrary $\gamma'\in\Gamma$.

Let $L$ be the set of primitive charges
such that the charge of an arbitrary
BPS particle is given by $n\gamma$ with
$\gamma\in L$ and $n\in\ZZ_+$.
The BPS spectrum
at a point of Coulomb branch
is encoded in a set of functions $K_\gamma(z)$
defined for each $\gamma\in L$.
These functions are called quantum Kontsevich-Soibelman factors.
The functional form of $K_\gamma(z)$ depends
on the helicity of the BPS particle $\gamma$.
(By an abuse of notation we use $\gamma$
as a label of sorts of particles.)
In the following we deal with only
BPS particles belonging to a half-hyper multiplet,
and the function $K_\gamma$ for such a particle is given by
\begin{align}
K_\gamma(z)=E_q(z)
=\Pexp\frac{q^{\frac{1}{2}}z}{1-q}
=\prod_{k=0}^\infty
\frac{1}{1-q^{\frac{1}{2}+k}z}
=\sum_{k=0}^\infty
\frac{q^{\frac{k}{2}}}{(q)_k}z^k,
\end{align}
where $(q)_k$ is the $q$-factorial defined by $(q)_k=\prod_{i=1}^k(1-q^i)$.

The Cordova and Shao's IR formula \cite{Cordova:2015nma} is
\begin{align}
I
&=\lambda^r\tr\left(
\prod^\curvearrowleft_{\gamma\in L} K_\gamma(X_\gamma)\right),
\end{align}
where 
$\prod^\curvearrowleft_{\gamma\in L}$
is the phase ordered product according to $\arg Z_\gamma$.
$r$ is the rank of the theory, which is the number of massless $U(1)$ gauge fields
at a generic point of the Coulomb branch.
$X_\gamma$ are operators satisfying
the quantum torus algebra
\footnote{
In the literature $q'$ defined by
$q'^{1/2}=-q^{1/2}$ is often used and then
the minus sign does not appear.}
\begin{align}
X_\gamma X_{\gamma'}
=(-q^{\frac{1}{2}})^{\langle\gamma,\gamma'\rangle}
X_{\gamma+\gamma'}
=q^{\langle\gamma,\gamma'\rangle}
X_{\gamma'}X_\gamma.
\label{xxalgebra}
\end{align}
The trace $\tr X_\gamma$ is defined as an
isomorphic map from $\Gamma_F$ to $\CC^*$.
If $\gamma\notin\Gamma_F$ $\tr X_\gamma=0$.

To physically interpret this formula,
we should understand the structure of
BPS configurations in the Coulomb branch \cite{Cordova:2016uwk}.
In general, a BPS configuration contains massive BPS particles
with different charges.
For a configuration in $\bm{S}^3\times\RR_t$ to preserve the supercharges $Q$ and $\ol Q$,
which are necessary to define the Schur index,
the particles must be aligned along a large circle in $\bm{S}^3$,
and the position $\theta_\gamma$ on the circle is
determined by the central charge $Z_\gamma$ by
$\theta_\gamma=\arg Z_\gamma$.
Thus we can specify the particle distribution
in a BPS configuration by giving a set of occupation numbers $n_\gamma$ for $\gamma\in L$.
We denote this set by $\{n_\gamma\}_{\gamma\in L}$.

If we
expand the function $K_\gamma$
as
\begin{align}
K_\gamma(z)=\sum_{n=0}^\infty C_\gamma(n)z^n,
\end{align}
then the index is given
as the summation over the occupation numbers:
\begin{align}
I
&=\sum_{\{n_\gamma\}_{\gamma\in L}}
\lambda^r
\left(\prod_{\gamma\in L} C_\gamma(n_\gamma)\right)
\tr\left(\prod^\curvearrowleft_{\gamma\in L} X_\gamma^{n_\gamma}\right).
\end{align}
The summand gives the contribution of a specific set of occupation numbers,
and
consists of the following three factors.
\begin{itemize}
\item
The factor $\lambda^r$ is the $1$-loop contribution of
the massless vector multiplets.

\item
The factor $\prod C$ is the contribution
associated with internal degrees of freedom of BPS particles.

\item
The trace factor can be identified with the
classical contribution of massless vector multiplets.
The angular momentum
induced by the Poynting vector due to the existence
of mutually nonlocal charges
contributes to the index by the factor
$e^{2\pi i(J+\ol J)}q^{J+\ol J}$,
and $q$-dependence of the trace factor gives
this factor.
For example, let us consider two adjucent charges $\gamma$ and $\gamma'$ on the
large circle.
If they carry mutually non-local charges, the induced massless gauge fields contribute to the
angular momentum $J+\ol J$ by $\pm\frac{1}{2}\langle\gamma,\gamma'\rangle$.
The angular momentum is independent of the distance between charges while the
signature depends on the order of the charges along the circle.
This order dependence is correctly reproduced by the algebra (\ref{xxalgebra}).
This factor also provides the dependence on the flavor fugacities.

\end{itemize}

Now we suggest an IR formula for the orbifold
as a natural generalization of the original one.
The $\ZZ_n$ acts on the large circle as a shift by $2\pi/n$
and the orbifolding makes it
a circle with circumference $2\pi/n$.
In other words, the large circle in the covering space consists of $n$
fundamental regions, and only BPS particles
in one of the fundamental regions are independent.
Let $L_n$ be the set of primitive charges associated with a specific
fundamental region.
The charge distribution of a BPS configuration in the orbifold
is specified by $\{n_\gamma\}_{\gamma\in L_n}$.
Therefore, the index should be given as the summation over
$\{n_\gamma\}_{\gamma\in L_n}$.

The first factor in the summand should be replaced by
the $U(1)$ factor $\lambda_n^r$ with $\lambda_n$ given in (\ref{lambdan}).
The second factor associated with the internal degrees of freedom
of BPS particles should be
the product of $C_\gamma(n_\gamma)$ for $\gamma\in L_n$.
The third factor, as we mentioned above, can be regarded as the classical contribution
to the angular momentum and the flavor charges.
The angular momentum can be obtained by integrating the contribution of
the Poynting vector over the orbifold.
The same result is obtained by first calculating the angular momentum for the covering space ${\bm S}^3$,
and dividing the result by $n$.
This is also the case for the flavor charges.
Correspondingly, the trace factor should be replaced by the $n$-th root of the trace for the configuration
in the covering space.
Combining these, we obtain
\begin{align}
I_n=
\sum_{\{n_\gamma\}_{L_n}}
\lambda_n^r
\left(\prod_{\gamma\in L_n}C_\gamma(n_\gamma)\right)
\left[\tr\left(\prod_{\gamma\in L}^\curvearrowleft X_\gamma^{n_\gamma}\right)\right]^{1/n}.
\label{irzn}
\end{align}
This is the IR formula we want to discuss in the next section.

An additional comment for the trace factor would be in order.
As we mentioned above, only charge distribution in a specific fundamental region is
independent, and the distribution in other regions should be determined by
the $\ZZ_n$ symmetry.
The phase ordered trance in (\ref{irzn}) must be taken over all charges in the
covering space.
It is important that the $\ZZ_n$ acts on charges non-trivially,
and the product is not simple $n$-th power of the product in the fundamental region.
In the $n=2$ case, for example, the $\ZZ_2$ action flips the signs of charges as we will explicitly see in the next section,
and the trace factor takes the form
$\left[\tr(X_{\gamma_1}^{n_1}X_{\gamma_2}^{n_2}\cdots X_{-\gamma_1}^{n_1}X_{-\gamma_2}^{n_2}\cdots)\right]^{1/2}$.

\section{Comparison}
Let us apply the IR formula
(\ref{irzn})
to a few simple examples
and compare the results with what are obtained by the UV formula.

We first consider the system of a free hypermultiplet $\vec q=(q,\wt q)$.
Before computing the index by the UV and IR formulae
it is important to check
the consistency between
BPS configurations in the Coulomb branch
and the orbifold action.

In the system of a single hypermultiplet,
we have only two sorts of particles with flavor charge $\pm\gamma$.
In a BPS configuration
there are particles with charge $+\gamma$ at $\theta_\gamma=\arg Z_\gamma$
and anti-particles with charge $-\gamma$ at $\theta_\gamma+\pi$.
This cannot be invariant under $\ZZ_n$ action
with $n\geq 3$.
Even for $n=2$ the $\ZZ_2$ exchanges
particles and anti-particles,
and the orbifolding seems to contradict the BPS configuration
in the Coulomb branch.
When $n=2$, actually, we can take it back to the original configuration
by performing additional charge conjugation.
This additional transformation is also necessary
to keep the mass term in the Lagrangian $\ZZ_2$
invariant.
To make the hypermultiplet massive we need to turn on
the vev $\langle\phi\rangle=m$ of the scalar component $\phi$ of the
non-dynamical background vector multiplet
coupling to the $U(1)_F$ current.
Because $\phi$ carries $U(1)_r$ charge $+2$,
the $g_2$ action
changes its sign.
This is compensated by the additional charge conjugation,
which flip the sign of the background vector multiplet.

The charge conjugation exchanging $q$ and $\wt q$
is realized by the $SU(2)_F$ transformation
$\vec q\rightarrow U\vec q$ with
\begin{align}
U=i\sigma_x.
\label{un2}
\end{align}
This anti-commutes with the $U(1)_F$ generator $i\sigma_z$,
and works as the charge conjugation.
For consistency, we should set the $U(1)_F$ Wilson line to be $z=\pm 1$.
(A generic Wilson line is not allowed because
the operator $z^F$ does not commute with $i\sigma_x$, and incompatible with the orbifold action.)

Now we have chosen the $\ZZ_2$ generator
$g_2U$
consistent to the Coulomb branch vev.
Let us calculate the index
by using UV and IR formulae.
When $z=\pm1$,
the orbifold with the charge conjugation twist
is in fact equivalent to
the $\ZZ_2$ orbifold with fractional holonomies studied in the previous section
up to $SU(2)_F$ rotation, because
the fractional holonomy $\omega_2^{hF}=\pm i\sigma_z$
is $SU(2)_F$ conjugate
to the charge conjugation
$U=i\sigma_x$.
Therefore, the index is still given by (\ref{hyperz2}) with $z=\pm1$.
By setting $z=1$, the one-particle index becomes
\begin{align}
i_2^{\pm\frac{1}{2}}=\frac{q^{\frac{1}{2}}}{1-q},
\label{irresult}
\end{align}
and the Schur index is
\begin{align}
I_2=\Pexp\frac{q^{\frac{1}{2}}}{1-q}=E_q(1).
\label{i2z}
\end{align}
On the IR side, we have only one
charge $\gamma$ in $L_2$.
The suggested formula (\ref{irzn}) gives
\begin{align}
I_2=\sum_{n=0}^\infty C_\gamma(n)=E_q(1),
\end{align}
and this agrees with (\ref{i2z}).

By introducing multiple hypermultiplets
we can realize more general orbifolds.
Let us consider the system of $k$ free hypermultiplets.
There are $2k$ sorts of particles.
For $\ZZ_{2k}$ invariance, we need to tune
the Coulomb branch vev $\phi$ and introduce
an appropriate twist $U_k$.
Let $\phi=\diag(a_1,\ldots,a_k)\otimes\sigma_z\in sp(k)$ be the Coulomb branch vev.
(We use the basis in which the $Sp(k)$ invariant tensor is given by $J=\bm{1}_k\otimes\epsilon$.)
We tune $a_i$ so that
the central charges are
$\ZZ_{2k}$ symmetric and given by
\begin{align}
Z(q_i)=a_i=\omega^i_{2k}a,\quad
Z(\wt q_i)=-a_i=\omega^{k+i}_{2k}a.
\end{align}
The action of $g_{2k}$ shifts $a_i$ to $a_{i+1}$ for $i=1\sim k-1$ and
$a_k$ to $-a_1$.
To keep the background unchanged,
we need a twist $U_k$ such that
the transformation
$\phi'=U_k\phi U_k^\dagger$ acts on $a_i$ as
\begin{align}
a_1'=-a_k,\quad
a_i'=a_{i-1}\quad(i=2\sim k).
\end{align}
The following $U_k$ realize this transformation:
\begin{align}
U_k=\left(\begin{array}{cccc}
& \bm{1}_2 \\
&& \ddots \\
&&& \bm{1}_2 \\
i\sigma_x
\end{array}\right)\in Sp(k).
\label{udef}
\end{align}
(This is a generalization of the charge conjugation (\ref{un2}) in the $k=1$ case.)
Just by the same reason as the $\ZZ_2$ case we set the Wilson line to vanish.
The diagonalization of $U_k$ gives the eigenvalues $\pm\alpha_m$ with
\begin{align}
\alpha_m=\exp\left(\frac{2\pi i}{2k}\left(m+\frac{1}{2}\right)\right),\quad
m=0,\ldots,k-1.
\end{align}
For each pair $(\alpha_m,-\alpha_m)$ of eigenvalues
this is the same as the action of the fractional holonomy $h=m+1/2$ on
a single hypermultiplet,
and the one-particle index
of the whole system
is the sum of (\ref{hyperzn}) over $h=1/2,\ldots,k-1/2$:
\begin{align}
i_n=\frac{q^{\frac{1}{2}}}{1-q^{2k}}(1+q+\cdots+q^{2k-1})=\frac{q^{\frac{1}{2}}}{1-q}.
\end{align}
The Schur index is $I_n=\Pexp i_n=E_q(1)$, and
this is the same as what we obtain from the IR formula.
Each fundamental region of the large circle contains one sort of
particles and
the IR formula
(\ref{irzn})
gives $I=E_q(1)$.

We can consider more general case in which
$n$ is not $2k$ but its divisor $n=2k/d$.
Such an orbifold is defined by using $(g_{2k}U_k)^d$ as the $\ZZ_n$ generator,
and
the one-particle index becomes
\begin{align}
i_n
=\frac{q^{\frac{1}{2}}}{1-q^k}\sum_{m=0}^{2k-1}q^{[m]_n}
=d\frac{q^{\frac{1}{2}}}{1-q},
\end{align}
and the Schur index is $I_n=\Pexp i_n=E_q(1)^d$.
Again, this is reproduced by the IR formula
(\ref{irzn})
by taking account of the
$d$ sorts of particles in a fundamental region.

Up to here we have found good agreement between UV and IR results.
Let us move on to a more complicated system,
QED with $N_f$ hypermultiplets.
Although this system is not conformal,
it is known that the two formulae give the same
answer for the ordinary Schur index \cite{Cordova:2015nma},
and it is natural to expect this holds for orbifolds, too.
However, disappointingly, we find discrepancy for orbifold index
in this case.

For example, let us consider $\ZZ_2$ orbifold of the $2$-flavored QED.
The UV formula gives the index (\ref{qednf2}).
When we use the IR formula we need to take account of the
existence of
two sorts of particles in the fundamental region.
Let $\gamma_1$ and $\gamma_2$ be their charges and
$n_1$ and $n_2$ be the corresponding occupation numbers.
In the covering space the charges of particles
in a fundamental region are always canceled by
the mirror images in the other fundamental region,
and the trace factor in (\ref{irzn}) does not impose any constraints
on $n_1$ and $n_2$.
If we assume the trivial flavor holonomy, the
trace factor simply gives $1$.
This means the index is essentially the same as that for two half-hyper multiplets
in $\bm{S}^3$, and given by
\begin{align}
I_2
=\lambda_2\sum_{n_1,n_2=0}^\infty C_\gamma(n_1)C_\gamma(n_2)
=\lambda_2E_q^2(1)
=1+2q^{\frac{1}{2}}+3q+6q^{\frac{3}{2}}+\cdots.
\end{align}
This is obviously different from the result
of the UV formula (\ref{qednf2}).

\section{Discussions}
In this paper we generalized the Schur index to the orbifold $\bm{S}^3/\ZZ_n$
in such a way that the $\ZZ_n$ action preserves the two supersymmetries
respected by the definition of the Schur index.
We also naively generalized the Cordova and Shao's IR formula for Schur index
to the orbifold case.
It reproduces the correct index for
a system of free hypermultiplets
when the fugacities and holonomies are
chosen in a $\ZZ_n$ invariant way.
However,
it is far from satisfactory.
It has the following deficits.
\begin{itemize}
\item
Although any SCFT admits $\ZZ_n$ orbifold with an arbitrary $n=2,3,\ldots$ in the UV description,
the IR formula works only for special values of $n$ depending on the theory.
\item
Even for a system of free hypermultiplets
it is not possible to turn on generic Wilson lines.

\item
For systems with dynamical vector multiplets
the formula does not reproduce the correct answer.

\end{itemize}

The second and third deficits may be related to each other.
Incompatibility of the generic Wilson lines seems to
prevent us from performing path integral
over all gauge field configurations.
At present, we cannot claim anything definite for this point,
and more investigation is desired.
It is important to study
more general systems
to clarify to what extent our formula
works and (if possible)
how we should improve it to make it
applicable to general systems.
In particular, it would be interesting and important to
study the orbifold Schur index of non-Lagrangian theories such as Argyres-Douglas theories.
Lagrangian theories
that flow to a class of Argyres-Douglus theories are proposed
in \cite{Maruyoshi:2016tqk,Maruyoshi:2016aim,Agarwal:2016pjo,Benvenuti:2017lle,Benvenuti:2017kud,Agarwal:2017roi,Benvenuti:2017bpg}
and this enable us to apply the UV formula of the orbifold index to the non-Lagrangian theories.

It is also interesting to analyze the relation between the orbifold index and chiral algebra.
In the unorbifolded case, the chiral algebra is realized
in a complex plane,
which is identified by Weyl rescaling with $\bm{S}^1\times\RR_t\subset \bm{S}^3\times\RR_t$, where
$\bm{S}^1$ is the large circle in $\bm{S}^3$ on which BPS particles are aligned.
The Virasoro generators $L_0$ and $\ol L_0$ on the plane are
related to the superconformal generators by
\begin{align}
L_0&=\frac{1}{2}(H+J+\ol J)=\frac{1}{2}(\mu_x+\mu_p)+\mu_q,\nonumber\\
\ol L_0&=\frac{1}{2}(H-J-\ol J-2R)=\frac{1}{2}(\mu_x+\mu_p).
\end{align}
For Schur operators with $\mu_x=\mu_p=0$
$\ol L_0=0$ and
the $\ZZ_n$ generator $g_n=\omega_n^{\mu_x-\mu_p+\mu_q}$ is given by
\begin{align}
g_n
=
\omega_n^{L_0}.
\end{align}
Namely, this is $\ZZ_n$ orbifold acting on the worldsheet.
Therefore, in the context of the chiral algebra, the orbifolding
should be regarded as the insertion of
a twist operator at the origin.
Such a relation is studied for an orbifold version of the Macdonald index in \cite{fluder}.
It would be interesting to do a similar analysis for the orbifold Schur index.

\section*{Acknowledgements}
I would like to thank Hirotaka Kato for valuable discussions.
This work was partially supported by Grand-in-Aid for Scientific Research (C) (No.15K05044),
Ministry of Education, Science and Culture, Japan.

\end{document}